\documentclass{naturearxiv}
\usepackage{graphicx}
\usepackage[space]{grffile}
\usepackage[nolists,nomarkers]{endfloat}
\usepackage{amsmath,amssymb}
\usepackage{mathptmx}

\newcommand{\f}{|f\rangle}
\newcommand{\e}{|e\rangle}
\newcommand{\fth}{|f;\,+\frac{3}{2}\rangle}
\renewcommand{\eth}{|e;\,+\frac{3}{2}\rangle}

\title{Evaporative cooling of the dipolar radical OH}

\author{Benjamin K. Stuhl$^1$, Matthew T. Hummon$^1$, Mark Yeo$^1$,
Goulven Qu{\'e}m{\'e}ner$^1$, John L. Bohn$^1$, \& Jun Ye$^1$}

\begin{document}

\maketitle

\begin{affiliations}
 \item JILA, National Institute of Standards and Technology and University of
Colorado, and Department of Physics, University of Colorado, Boulder, CO 80309, USA.
\end{affiliations}

\spacing{1}
\begin{abstract}
Atomic physics was revolutionized by the development of forced evaporative cooling:
it led directly to the observation of Bose-Einstein condensation\cite{Anderson14071995,PhysRevLett.75.3969}, quantum-degenerate
Fermi gases\cite{DeMarco10091999}, and ultracold optical lattice simulations
of condensed matter phenomena\cite{Bakr30072010}. More recently, great progress
has been made in the production of cold molecular gases\cite{Ni2008}, whose permanent electric
dipole moment is expected to generate rich, novel, and controllable phases\cite{Pupillo2008,PhysRevA.83.043602,PhysRevA.84.013603},
dynamics\cite{PhysRevLett.96.190401,PhysRevLett.98.060404,PhysRevLett.107.115301}, and chemistry\cite{Ospelkaus2010,Ni2010,Quemener2012} in these
ultracold systems. However, while many strides have been made\cite{Carr:2009oz} in both direct cooling
and cold-association techniques, evaporative cooling has not yet been achieved
due to unfavorable elastic-to-inelastic ratios\cite{Ni2010} and impractically slow thermalization
rates in the available trapped species. We now report the observation of
microwave-forced evaporative cooling of hydroxyl (OH) molecules
loaded from a Stark-decelerated beam into an extremely high-gradient magnetic quadrupole trap.
We demonstrate cooling by at least an order of magnitude in temperature
and three orders in phase-space density, limited only by the low-temperature
sensitivity of our spectroscopic thermometry technique. With evaporative cooling
and sufficiently large initial populations, much colder temperatures are possible,
and even a quantum-degenerate gas of this dipolar radical -- or anything else
it can sympathetically cool -- may now be in reach.
\end{abstract}

Evaporative cooling of a thermal distribution\cite{Ketterle:1996bv} is, in principle, very simple:
by selectively removing particles with much greater than the average total energy
per particle, the temperature decreases. In the presence of
elastic collisions, the high-energy tail is repopulated and so may \emph{repeatedly} be selectively
trimmed, allowing the removal of a great deal of energy at low
cost in particle number. This process may be started as soon as the thermalization
rate is fast enough to be practical and continued until its cooling power
is balanced by the heating rate from inelastic collisions. It generally yields
temperatures deep into the quantum-degenerate regime (far below the recoil limit of optical cooling).

The key metric for evaporation is therefore the ratio of two timescales. The first
is the rate of elastic collisions, which rethermalize the distribution, while the second is the rate at which particles are lost
from the trap for reasons other than their being deliberately removed, e.~g. the
rates of inelastic scattering and background gas collisions. Both theoretical\cite{Lara_Rb_OH,PhysRevA.79.062708,Quemener2012}
and experimental work\cite{campbell2009mechanism,Ni2010,PhysRevLett.106.193201} have seemed to show a generically poor value of
this ratio across multiple molecular systems; this has led to a general belief that evaporative cooling is unfavorable
in molecules\cite{Carr:2009oz}. As no trapped molecular system has achieved sufficiently rapid
thermalization, there has been a lack of experiments to test this expectation.

Hydroxyl would not, at first glance, seem to be a promising candidate for
evaporative cooling. Its open-shell ${^2}\Pi_{3/2}$ ground state and its
propensity towards hydrogen bonding create a large anisotropy in the OH--OH
interaction potential, which would intuitively motivate a large inelastic scattering rate. Chemical reactions are also possible, via
the OH$+$OH$\rightarrow$H$_2$O$+$O pathway; it is unclear whether this reaction
has an activation energy barrier\cite{acp-4-1461-2004,doi:10.1021/jp066359c}.
% from Goulven, though modified, here through "<---"
% -->
It is thus perhaps surprising that the elastic collision rate actually exceeds the inelastic
rate, allowing evaporative cooling. However, our experimental observation is unambiguous,
and is further supported by quantum scattering
calculations based on the long range dipole-dipole interaction between the
molecules\cite{Avdeenkov2002,Ticknor2005} considering all of the
fine-structure states of the rotational ground state.  This analysis considers only
elastic collision or inelastic relaxation to lower energy states, as the long-range
interactions appear to fully dominate over short-range effects such as chemical
reactions.

\begin{figure}
\includegraphics{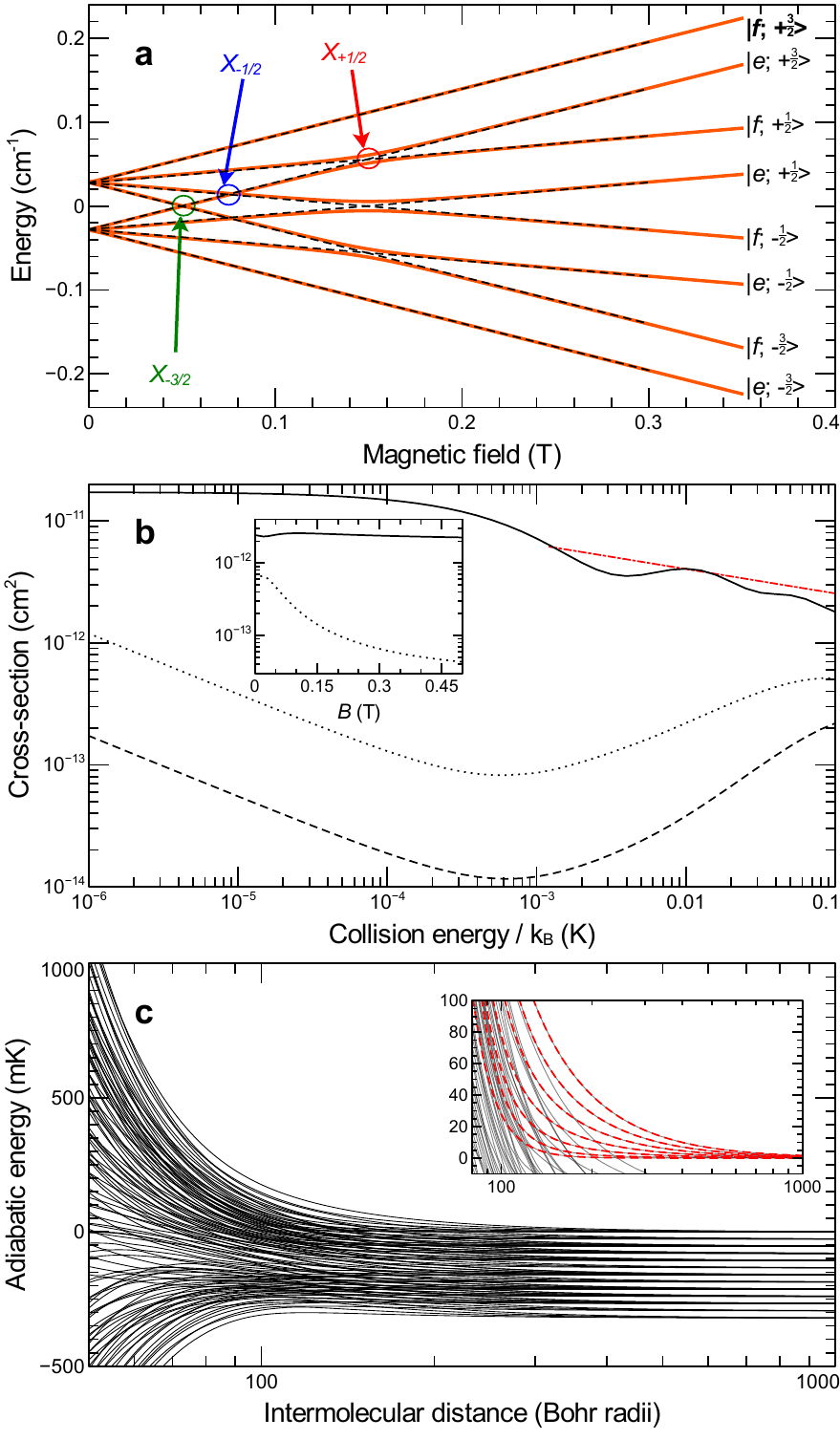}
\caption{\label{fig:theory}
\textbf{Ground-state structure and scattering theory of OH.}
\textbf{a},~The ground-state $\Lambda$-doublet and Zeeman structure of OH. Dashed
black lines are energy levels in the absence of any electric field. In the presence of
an electric field, the level crossings become avoided (solid orange lines).
$X_i$ label the crossings of the $\eth$ state with the $|f; \, M_J = i \rangle$ states;
these crossings allow $\eth$ molecules to escape the trap in the presence of an electric field.
Molecules are loaded into the magnetic trap in the bolded $\fth$ state.
\textbf{b},~Cross sections as a function of collision energy: elastic cross section
in a 50~mT magnetic field (solid), inelastic in 50~mT (dotted), and inelastic in 150~mT (dashed).
The red dash-dotted line is a semi-classical expression given by Eq.~\eqref{eqn:Child}.
Inset: elastic (solid) and inelastic (dotted) cross sections as a function of magnetic field at $E_c=50$~mK.
\textbf{c},~Adiabatic energies as a function of the inter-molecular distance at $B=50$~mT. The inset
zooms in on the repulsive van der Waals interaction for the case of two colliding
$\fth$ molecules.
}
\end{figure}

In its ground state, OH has a $^2\Pi_{3/2}$ electronic character, with the
lowest rotational level having total non-nuclear angular momentum $J=\frac{3}{2}$. The electronic orbital angular momentum couples to the rotational angular
momentum to split the two opposite-parity states within $J=\frac{3}{2}$ by a
$\Lambda$-doubling of $\sim\!1.667$~GHz; the upper parity state is labeled $\f$ and the
lower $\e$. Hydroxyl is both paramagnetic, with a molecule-fixed moment of 2$\mu_B$ ($\mu_B$ is the Bohr magneton),
and electrically polar with a dipole moment of 1.67~Debye ($5.57 \times 10^{-30}$~C$\cdot$m). The Zeeman spectrum
of OH is shown in Fig.~\ref{fig:theory}a. Our magnetic trap\cite{Sawyer2008a} is loaded with
molecules in the uppermost $|f;\,M_{J}=+\frac{3}{2}\rangle$ state, where $M_J$ is the
laboratory projection of $J$.

The results of our scattering calculations for $\fth$ molecules
are shown in Fig.~\ref{fig:theory}b.
The elastic cross section dominates the inelastic one for low energies:
at a collision energy of $E_c=50$~mK, the ratio of
elastic over inelastic is $R=5$ in a 50~mT magnetic field ($B$) and $R=18$ in 150~mT, while at lower energy $E_c=5$~mK the ratio
increases to $R=23$ and $137$ respectively. Because the collisions occur in a quadrupole magnetic
trap where $B$ is inhomogeneous, the inset
of Fig.~\ref{fig:theory}b shows the cross sections as a function of $B$
at $E_c=50$~mK. This demonstrates that inelastic processes are even further
suppressed at $B > 50$~mT, in agreement with previous analysis~\cite{Ticknor2005}.

These scattering results can be interpreted by the emergence of an effective
\emph{repulsive} van der Waals interaction between the two molecules.  In zero electric
field, the effective interaction in a scattering channel $m$ can be evaluated in second-order perturbation theory by
\begin{eqnarray}
V_{\text{vdW}}(r) = \sum_{n} \frac{ | \langle  m | V_{dd}(r) |  n \rangle |^2 }
{ E_m - E_n } \sim \frac{ C_6 }{ r^6 },
\label{eqn:C6_def}
\end{eqnarray}
where $V_{dd}$ is the electric dipole-dipole interaction between different scattering
channels, and is non-vanishing only between molecular states of distinct parity;
and $E_m$ and $E_n$ are the asymptotic energies of the relevant
scattering channels.  For the initial molecule-molecule channel of interest
$\fth \, \fth$, the other fine-structure channels are lower in
energy ($E_n < E_m$) and repel this energy upward.  Thus, the $C_6$ coefficient is positive, as illustrated in Fig.~\ref{fig:theory}c for $B=50$~mT,
for the highest energy channel. The contribution from the next rotational
state $J=\frac{5}{2}$ of the $^2\Pi_{3/2}$ manifold is too high in energy ($E_m - E_{J=5/2} \approx -100$~K)
to give an appreciable attractive contribution at long-range. Hence, for
low collision energies, the scattering of OH molecules is dominated by the long-range interaction rather
than the short-range structure of the potential surface.

Given an effective, repulsive $C_6$ coefficient in the incident channel,
the elastic cross section can be approximated semi-classically at energies
above the threshold regime\cite{Child_Book_1996}, as
\begin{eqnarray}\label{eqn:Child}
\sigma^{\text{el}} & = & \frac{\pi^{11/5} \, (\Gamma(5/2)/\Gamma(3))^{2/5}}{\sin{(\pi/5)} \, \Gamma(2/5)} \, \bigg(\frac{\bar{C}_6}{\hbar}\bigg)^{2/5} \, \bigg(\frac{2 E_c}{m_{\text{red}}}\bigg)^{-1/5}
\end{eqnarray}
where $m_{\text{red}}=14497$ atomic units (a.~u.) is the reduced mass of two OH molecules and
$\bar{C}_6 \approx 9 \times 10^4$~a.~u. is the calculated isotropic $C_6$.
(We contrast this $\bar{C}_6$ with values of 2--8$\times 10^3$~a.~u.
for the alkali metal atoms\cite{PhysRevLett.82.3589}.)  Equation~\eqref{eqn:Child}, plotted as
a red dash-dotted line on Fig.~\ref{fig:theory}b, only slightly overestimates the numerical results
for the elastic cross section but gives a proper trend in $E_c^{-1/5}$.
% <-- (end of material from Goulven)

\begin{figure}
\includegraphics{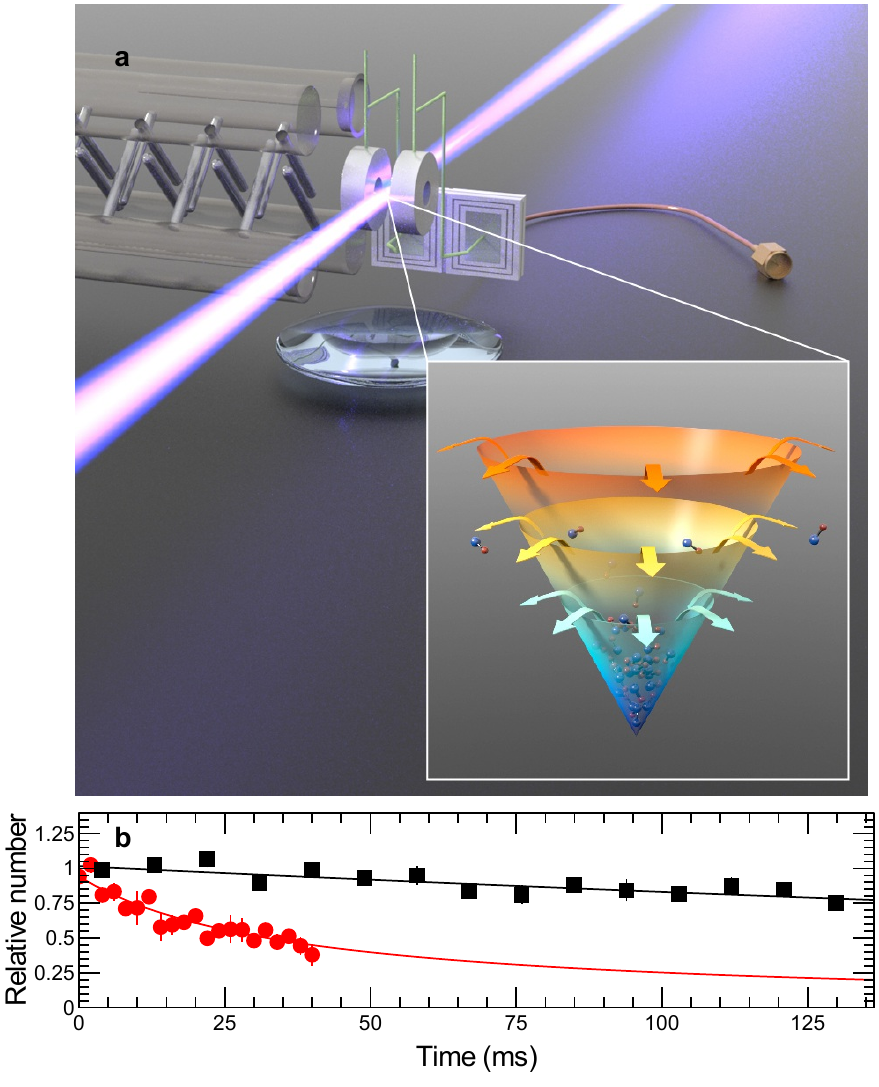}
\caption{\label{fig:cartoon}
\textbf{Trap system and inelastic collisions.}
\textbf{a}, Schematic of the Stark decelerator\cite{Bochinski2003} (left) and magnetic trap (center rings) system.
The DC-blocking capacitor (rear) decouples the high voltage used for trap loading from
the microwave system used for spectroscopy and evaporation, while the laser and lens
provide fluorescence detection of the trapped molecules.
Inset: an artist's impression of evaporative cooling. As the trap depth is successively
lowered by the RF knife, the hottest molecules escape and the remainder rethermalize to a colder temperature.
\textbf{b}, Time-of-flight trace of electric field-induced collisions at 45~mK. Black squares,
consistent with pure one-body loss, are with no applied electric field; red circles
are well-fit by pure two-body inelastic loss induced by a large applied electric field of 3040~V/cm.
(The field used in the RF knife is only 240~V/cm, for comparison.) Error bars are one standard error.
}
\end{figure}

We observed inelastic scattering in the presence of a large
electric field (Fig.~\ref{fig:cartoon}b), demonstrating the presence
of two-body collisions in our trapped sample. Motivated by the prediction of a favorable
$R$, we then undertook an experiment
to implement microwave-forced evaporative cooling. The $\fth \rightarrow \eth$
microwave transition has a small but nonzero differential Zeeman shift\cite{PhysRevA.74.061402}, red-shifting
by 26.6~kHz/mT. In the presence of a small electric field, $\eth$ molecules can escape the
trap through the avoided crossings labeled $X_i$ in Fig.~\ref{fig:theory}a\cite{PhysRevA.85.033427}, while inelastic
losses of $\f$ molecules remain unmeasurably slow.
A brief microwave pulse to selectively transfer $\fth$-state molecules to $\eth$,
followed by a longer period with an electric field present to eject $\eth$
molecules from the trap, is therefore a field- (position-)
selective method to remove molecules from the magnetic trap (see Methods).
This yields what is commonly called an RF knife (Fig.~\ref{fig:cartoon}a).
A Zeeman depletion spectrum can be acquired by using the knife to remove molecules
at a set of fixed frequencies and measuring the fractional depletion $\zeta$; this yields a histogram
of relative molecule number versus $B$. We fit this spectrum with a modified thermal distribution
\begin{align}
\label{eqn:Boltzmann}
\zeta(B) dB & = \zeta_{0} B^2 dB \times \exp[-\frac{\mu B}{k_{B} T}] \times
\nonumber\\ & \qquad \times
  \begin{cases}
    1, & \mbox{if } B > 49.6 \mbox{~mT} \\
    \exp\{-\frac{\mu [(49.6 \;\mathrm{mT}) - B]}{k_{B} T}\}, & \mbox{otherwise}
  \end{cases}
\end{align}
where $\zeta_{0}$ is a fitting coefficient,
$B^{2} dB \propto r^2 dr$ is the volume element for a 3-D quadrupole trap,
$\mu = 1.2$~$\mu_B$ is the magnetic moment of the $\fth$ state, $B$ is the magnetic field
strength implied by the microwave frequency, $k_{B}$ is Boltzmann's constant,
and $T$ is the fitted temperature. The first term in the product is the simple
Boltzmann expression for the molecule number as a function of $B$, while the
second is a correction for the fact that $\eth$ molecules only disappear when they
go through one of the $X_i$ crossings. Specifically, while molecules transferred
at fields above 49.6~mT (the known location of $X_{-3/2}$) are always energetically
able to reach one of the $X_i$ crossings and thus disappear, of the molecules
transferred at lower fields only those with enough kinetic energy to fly up the trap potential to
$X_{-3/2}$ can escape the trap. This implies an additional Boltzmann factor $\exp[-\mu \Delta B/k_{B} T]$
in the probability of those molecules' disappearance.

\begin{figure}
\includegraphics{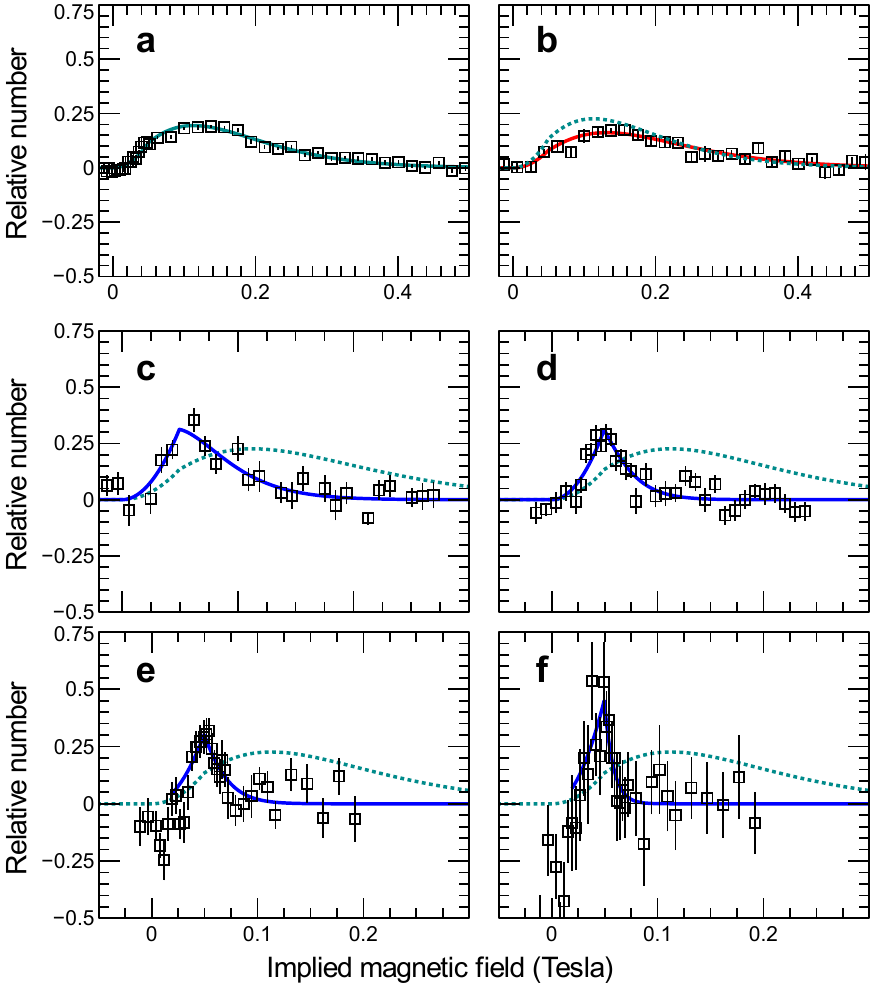}
\caption{\label{fig:spectra}
\textbf{Microwave spectra illustrating six different final temperatures.} Black squares
are data (error bars are 1 standard error), and solid lines are fits to the
sensitivity-corrected Boltzmann form of Eq.~\eqref{eqn:Boltzmann}: \textbf{a} the
unperturbed 45~mK distribution, \textbf{b} \emph{anti}-evaporation to 54~mK, \textbf{c}-\textbf{f}
forward evaporation to \textbf{c} 18~mK, \textbf{d} 12~mK, \textbf{e} 9.8~mK,
and \textbf{f} 5.1~mK. Dotted lines reproduce the fit from \textbf{a},
shown for comparison. Note that the x-axis scale differs between \textbf{a}-\textbf{b}
and \textbf{c}-\textbf{f}; all curves have been shifted vertically to have a zero baseline.
}
\end{figure}

With a $B$-selective technique for removing molecules, we easily
implement forced evaporation by moving the edge of the RF
knife from some large initial value of $B$ towards zero, at a rate slow enough
that the distribution remains in thermal equilibrium. Six different Zeeman spectra are shown in
Fig.~\ref{fig:spectra}, demonstrating both forced evaporation and forced
\emph{anti}-evaporation, where for the latter the knife is held fixed at some low $B_{\mathrm{knife}}$
($\mu B_{\mathrm{knife}} < k_{B} T$).
The trap is initially loaded with a temperature of 51~mK; left unperturbed, it
free-evaporates down to 45~mK. We have forced evaporative cooling by an order
of magnitude down to 5.1~mK, while forced anti-evaporation can overcome the free evaporation
and increase the temperature to 54~mK. The limit of 5.1~mK is attained approximately when the
RF knife edge reaches $X_{-3/2}$: while further reductions in temperature are possible,
the exponential suppression of the spectroscopic signal below
$X_{-3/2}$ renders our current system unable to \emph{measure} temperatures lower
than this. The plots of Fig.~\ref{fig:spectra} are all normalized so that the area under the 
spectroscopic curve is proportional to the total $\f$-state fluorescence signal. Thus, the increase in
signal height at low $B$ in Fig.~\ref{fig:spectra}c-f is direct evidence
of increasing phase-space density.

The apparent \emph{negative} signal in Fig.~\ref{fig:spectra}e-f
can be fully fitted by assuming the presence of accumulated, trapped $\e$-state molecules
in thermal equilibrium with the visible $\f$-state ones. Since $\eth$ molecules
are totally trapped if they do not have enough kinetic energy to reach $X_{-3/2}$, they
will \emph{appear}, rather than disappear, during the microwave spectroscopy
and contribute to the total depletion signal with a negative sign. Fitting the curves in this fashion
gives even colder temperatures of 6.8 and 3.5 mK for Figs.~\ref{fig:spectra}e and f,
respectively. As the appearance of low-energy, trapped $\e$-state molecules would
also constitute direct evidence of evaporation, we undertook a direct search and indeed detected
them in laser-induced fluorescence.

\begin{figure}
\includegraphics{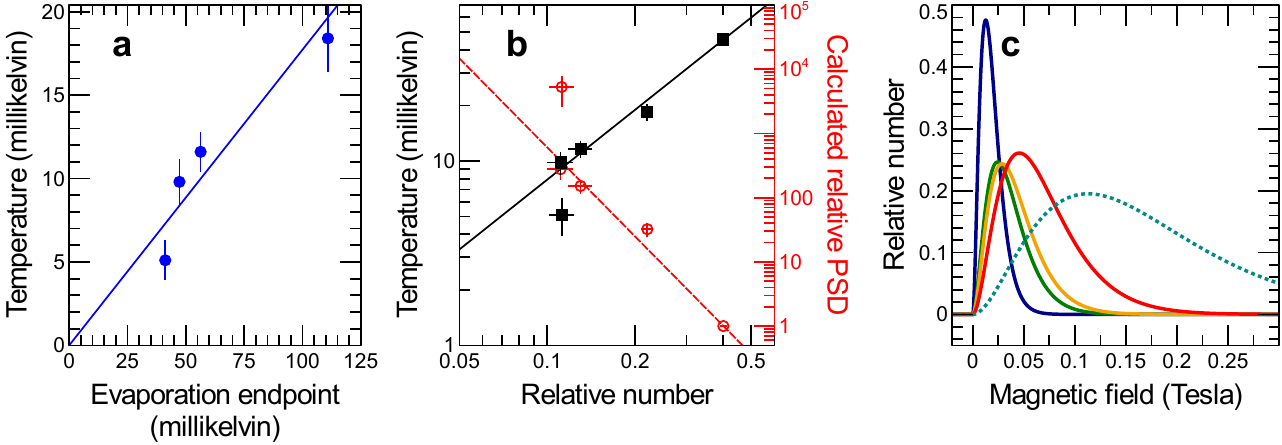}
\caption{\label{fig:PSD}
\textbf{Scaling relations observed in the evaporative cooling of OH.}
\textbf{a},~Final observed temperature versus effective trap depth at evaporation ramp end.
\textbf{b},~Final temperature (solid black, left axis) and calculated relative
phase-space density (dashed red, right axis) versus remaining relative
molecule number. Error bars are 1 standard error; phase space density is calculated
assuming a 3-D quadrupole trap geometry.
\textbf{c},~Molecular Boltzmann distributions implied by the fit curves
of Fig.~\ref{fig:spectra}a and c-f (right to left). Distributions are normalized so that
the area under each curve is proportional to the total $\f$-state fluorescence observed
at that temperature.
}
\end{figure}

We make several comments on the observed evaporation. The first is on the evaporation
timescale: it is \emph{fast}. In comparison to typical cooling rates of $d(\log T)/dt \sim 0.5$~s$^{-1}$,
we cool by an order of magnitude in only 70~ms, for a rate of 33~s$^{-1}$. This implies
elastic collision rates on the order of 100--1000~s$^{-1}$, comparable to our trap
frequency\cite{PhysRevA.85.033427} of $\sim\!1400$~s$^{-1}$. As we are able to set
a bound on the inelastic loss rate of $< 2$~s$^{-1}$ (as shown in Fig.~\ref{fig:cartoon}b),
this implies a distribution-averaged $R \gtrsim$ 50-500, consistent with the $B$-field dependence
of $R$ shown in Fig.~\ref{fig:theory}b.

We quantify the behavior of the evaporation by a set of power-law
scaling relations\cite{Ketterle:1996bv}, three of which are plotted in Fig.~\ref{fig:PSD}.
The average energy removed per molecule, $\eta \equiv \mbox{(trap depth)}/k_{B}T$,
is the first metric of interest: we observe a ratio $\eta = 5.6$, as shown in Fig.~\ref{fig:PSD}a.
(For scale, $\eta = $5--10 is considered reasonable in atomic evaporation\cite{Ketterle:1996bv}.)
Such a relatively low value of $\eta$ is unsurprising, given both the possible low ($< 1000$)
value of the elastic-to-inelastic ratio $R$ and the fact that molecules are only
actually lost when they cross the spatial regions corresponding to the $X_{i}$
crossings. This reduces the selectivity of the RF knife, as molecules transferred
to the $\e$-state may recollide and rethermalize before they find their way to
a crossing. The other metrics are the behavior of temperature and
relative phase-space density (PSD) as function of remaining molecule number, shown in Fig.~\ref{fig:PSD}b.
The efficiency of evaporation is determined by the number of molecules
remaining at a given temperature and PSD, that is by $\alpha \equiv d(\log T)/d(\log N)$
and $\gamma \equiv -d(\log \mathrm{PSD})/d(\log N)$. We measure $\alpha = 1.26$
and so using the fact that density scales as $\frac{1}{T^3}$ for our 3-D quadrupole trap
we find $\gamma = 4.7$. Extrapolating this $\gamma$ indicates that it would take
roughly a factor of 200 reduction in number to increase PSD by $10^{10}$.

We calculate both $\alpha$ and $\gamma$ assuming that our fluorescence signal
is linear with total molecule number, justified by the optical power broadening being larger
than the Zeeman broadening of the trap.
It is very difficult to determine the
sensitivity of pulsed-laser-induced fluorescence \textit{in situ}, so we use the
observed collision rate to estimate a lower bound on the density in our trap:
assuming a unitarity-limited scattering rate $\beta_\text{u}$ with a maximum
collisional angular momentum of $12\hbar$ (and the elastic scattering rate in
Fig.~\ref{fig:theory}b is only a factor of 3 below this value), an elastic collision
rate of $\beta_\text{u} n_0 = 300$~s$^{-1}$ implies a peak density $n_{0} \approx 5 \times 10^{10}$~cm$^{-3}$.
(Integration of a Boltzmann distribution with this peak density over the trap suggests a total
number of $\sim \! 10^6$ molecules in the free-evaporated trap sample and an initial peak PSD of $3 \times 10^{-10}$.)
This density is high enough to permit the use of absorption imaging
techniques to directly visualize the trap distribution in the future. Since imaging
allows direct, non-spectroscopic measurement of both density and temperature,
it will enable us to quantify further reductions in temperature. With the addition
of a mechanism to prevent Majorana loss\cite{Ketterle:1996bv}, the favorability
of $R$ down to microkelvin temperatures indicates that even Bose-Einstein
condensation of OH may be feasible.

\begin{methodssummary}
Our Stark decelerator and permanent magnet trap have been described elsewhere\cite{Bochinski2003,Sawyer2008a}
and are illustrated in Fig.~\ref{fig:cartoon}a. Briefly,
OH molecules are formed by an electric discharge through a saturated mixture of water
vapor in 150 kPa of krypton, supersonically expanded through a pulsed valve. The gas
packet is skimmed, focused by an electrostatic hexapole, and Stark-decelerated to 34~m/s.
The slowed packet is then stopped at the center of the magnetic quadrupole trap by
a high-voltage field applied between the permanent magnets, and thus loaded into the
magnetic trap.

Once trapped, the molecules are allowed to settle briefly (on the order of 5~ms) before
evaporation begins. The evaporation sequence consists of alternating microwave
(80~$\mu$s with 0~dBm at the vacuum feedthrough) and DC bias field
($\sim\!650$~$\mu$s at 240~V/cm) pulses: the microwave pulse selectively transfers hot
molecules from the $\f$-state to the $\e$-state, while
the DC bias destabilizes the $\e$-state so that those molecules are lost from the trap.
The microwave frequency is ramped along an exponential curve towards zero trap depth, truncated
at the desired final depth.

After evaporation, the spectroscopy sequence is executed. It is similar to the evaporation,
except that rather than slowly ramping a set of quasi-single-frequency pulses, each pulse
is rapidly swept through the same narrow (50-500~kHz) frequency band with an additional sine-wave
amplitude modulation (+10~dBm peak power at the feedthrough)
so as to induce Adiabatic Rapid Passage (ARP) transference\cite{PhysRevA.85.033427} of all
the $\f$-state molecules within the frequency band to the $\e$-state. A DC bias field again
rejects the $\e$-state molecules from the trap. The number of ARP pulses applied is generally
between 5 and 30, over a 2-10~ms spectroscopy sequence. The final molecule number is then
detected by pulsed laser-induced fluorescence using the 282~nm
$X{^2}\Pi_{3/2},v''=0\;{\rightarrow}\;A{^2}\Sigma,v'=1$ transition.

\end{methodssummary}

%% Put the bibliography here, most people will use BiBTeX in
%% which case the environment below should be replaced with
%% the \bibliography{} command.

% \begin{thebibliography}{1}
% \bibitem{dummy} Articles are restricted to 50 references, Letters
% to 30.
% \bibitem{dummyb} No compound references -- only one source per
% reference.
% \end{thebibliography}

%\bibliography{../references}

\begin{thebibliography}{10}
\expandafter\ifx\csname url\endcsname\relax
  \def\url#1{\texttt{#1}}\fi
\expandafter\ifx\csname urlprefix\endcsname\relax\def\urlprefix{URL }\fi
\providecommand{\bibinfo}[2]{#2}
\providecommand{\eprint}[2][]{\url{#2}}

\bibitem{Anderson14071995}
\bibinfo{author}{Anderson, M.~H.}, \bibinfo{author}{Ensher, J.~R.},
  \bibinfo{author}{Matthews, M.~R.}, \bibinfo{author}{Wieman, C.~E.} \&
  \bibinfo{author}{Cornell, E.~A.}
\newblock \bibinfo{title}{Observation of {Bose}-{Einstein} condensation in a
  dilute atomic vapor}.
\newblock \emph{\bibinfo{journal}{Science}} \textbf{\bibinfo{volume}{269}},
  \bibinfo{pages}{198--201} (\bibinfo{year}{1995}).

\bibitem{PhysRevLett.75.3969}
\bibinfo{author}{Davis, K.~B.} \emph{et~al.}
\newblock \bibinfo{title}{{Bose}-{Einstein} condensation in a gas of sodium
  atoms}.
\newblock \emph{\bibinfo{journal}{Phys. Rev. Lett.}}
  \textbf{\bibinfo{volume}{75}}, \bibinfo{pages}{3969--3973}
  (\bibinfo{year}{1995}).

\bibitem{DeMarco10091999}
\bibinfo{author}{DeMarco, B.} \& \bibinfo{author}{Jin, D.~S.}
\newblock \bibinfo{title}{Onset of {Fermi} degeneracy in a trapped atomic gas}.
\newblock \emph{\bibinfo{journal}{Science}} \textbf{\bibinfo{volume}{285}},
  \bibinfo{pages}{1703--1706} (\bibinfo{year}{1999}).

\bibitem{Bakr30072010}
\bibinfo{author}{Bakr, W.~S.} \emph{et~al.}
\newblock \bibinfo{title}{Probing the superfluid-to-{Mott} insulator transition
  at the single-atom level}.
\newblock \emph{\bibinfo{journal}{Science}} \textbf{\bibinfo{volume}{329}},
  \bibinfo{pages}{547--550} (\bibinfo{year}{2010}).

\bibitem{Ni2008}
\bibinfo{author}{Ni, K.-K.} \emph{et~al.}
\newblock \bibinfo{title}{A high phase-space-density gas of polar molecules}.
\newblock \emph{\bibinfo{journal}{Science}} \textbf{\bibinfo{volume}{322}},
  \bibinfo{pages}{231--235} (\bibinfo{year}{2008}).

\bibitem{Pupillo2008}
\bibinfo{author}{Pupillo, G.} \emph{et~al.}
\newblock \bibinfo{title}{Cold atoms and molecules in self-assembled dipolar
  lattices}.
\newblock \emph{\bibinfo{journal}{Phys. Rev. Lett.}}
  \textbf{\bibinfo{volume}{100}}, \bibinfo{pages}{050402}
  (\bibinfo{year}{2008}).

\bibitem{PhysRevA.83.043602}
\bibinfo{author}{Baranov, M.~A.}, \bibinfo{author}{Micheli, A.},
  \bibinfo{author}{Ronen, S.} \& \bibinfo{author}{Zoller, P.}
\newblock \bibinfo{title}{Bilayer superfluidity of fermionic polar molecules:
  Many-body effects}.
\newblock \emph{\bibinfo{journal}{Phys. Rev. A}} \textbf{\bibinfo{volume}{83}},
  \bibinfo{pages}{043602} (\bibinfo{year}{2011}).

\bibitem{PhysRevA.84.013603}
\bibinfo{author}{Levinsen, J.}, \bibinfo{author}{Cooper, N.~R.} \&
  \bibinfo{author}{Shlyapnikov, G.~V.}
\newblock \bibinfo{title}{Topological ${p}_{x}+{\mathit{ip}}_{y}$ superfluid
  phase of fermionic polar molecules}.
\newblock \emph{\bibinfo{journal}{Phys. Rev. A}} \textbf{\bibinfo{volume}{84}},
  \bibinfo{pages}{013603} (\bibinfo{year}{2011}).

\bibitem{PhysRevLett.96.190401}
\bibinfo{author}{Barnett, R.}, \bibinfo{author}{Petrov, D.},
  \bibinfo{author}{Lukin, M.} \& \bibinfo{author}{Demler, E.}
\newblock \bibinfo{title}{Quantum magnetism with multicomponent dipolar
  molecules in an optical lattice}.
\newblock \emph{\bibinfo{journal}{Phys. Rev. Lett.}}
  \textbf{\bibinfo{volume}{96}}, \bibinfo{pages}{190401}
  (\bibinfo{year}{2006}).

\bibitem{PhysRevLett.98.060404}
\bibinfo{author}{B\"uchler, H.~P.} \emph{et~al.}
\newblock \bibinfo{title}{Strongly correlated {2D} quantum phases with cold
  polar molecules: Controlling the shape of the interaction potential}.
\newblock \emph{\bibinfo{journal}{Phys. Rev. Lett.}}
  \textbf{\bibinfo{volume}{98}}, \bibinfo{pages}{060404}
  (\bibinfo{year}{2007}).

\bibitem{PhysRevLett.107.115301}
\bibinfo{author}{Gorshkov, A.~V.} \emph{et~al.}
\newblock \bibinfo{title}{Tunable superfluidity and quantum magnetism with
  ultracold polar molecules}.
\newblock \emph{\bibinfo{journal}{Phys. Rev. Lett.}}
  \textbf{\bibinfo{volume}{107}}, \bibinfo{pages}{115301}
  (\bibinfo{year}{2011}).

\bibitem{Ospelkaus2010}
\bibinfo{author}{Ospelkaus, S.} \emph{et~al.}
\newblock \bibinfo{title}{Quantum-state controlled chemical reactions of
  ultracold potassium-rubidium molecules}.
\newblock \emph{\bibinfo{journal}{Science}} \textbf{\bibinfo{volume}{327}},
  \bibinfo{pages}{853--857} (\bibinfo{year}{2010}).

\bibitem{Ni2010}
\bibinfo{author}{Ni, K.-K.} \emph{et~al.}
\newblock \bibinfo{title}{Dipolar collisions of polar molecules in the quantum
  regime}.
\newblock \emph{\bibinfo{journal}{Nature}} \textbf{\bibinfo{volume}{464}},
  \bibinfo{pages}{1324--1328} (\bibinfo{year}{2010}).

\bibitem{Quemener2012}
\bibinfo{author}{Qu{\'e}m{\'e}ner, G.} \& \bibinfo{author}{Julienne, P.~S.}
\newblock \bibinfo{title}{Ultracold molecules under control!}
\newblock \emph{\bibinfo{journal}{Chemical Reviews}}  (\bibinfo{year}{in
  press})
\newblock DOI: \bibinfo{doi}{10.1021/cr300092g}.

\bibitem{Carr:2009oz}
\bibinfo{author}{Carr, L.~D.}, \bibinfo{author}{DeMille, D.},
  \bibinfo{author}{Krems, R.~V.} \& \bibinfo{author}{Ye, J.}
\newblock \bibinfo{title}{Cold and ultracold molecules: science, technology and
  applications}.
\newblock \emph{\bibinfo{journal}{New J. Phys.}} \textbf{\bibinfo{volume}{11}},
  \bibinfo{pages}{055049} (\bibinfo{year}{2009}).

\bibitem{Ketterle:1996bv}
\bibinfo{author}{Ketterle, W.} \& \bibinfo{author}{VanDruten, N.}
\newblock \bibinfo{title}{Evaporative cooling of trapped atoms}.
\newblock \emph{\bibinfo{journal}{Adv. At. Mo. Opt. Phys.}}
  \textbf{\bibinfo{volume}{37}}, \bibinfo{pages}{181--236}
  (\bibinfo{year}{1996}).

\bibitem{Lara_Rb_OH}
\bibinfo{author}{Lara, M.}, \bibinfo{author}{Bohn, J.~L.},
  \bibinfo{author}{Potter, D.~E.}, \bibinfo{author}{Soldan, P.} \&
  \bibinfo{author}{Hutson, J.~M.}
\newblock \bibinfo{title}{Ultracold {Rb}--{OH} collisions and prospects for
  sympathetic cooling}.
\newblock \emph{\bibinfo{journal}{Phys.\ Rev.\ Lett.}}
  \textbf{\bibinfo{volume}{97}}, \bibinfo{pages}{183201}
  (\bibinfo{year}{2006}).

\bibitem{PhysRevA.79.062708}
\bibinfo{author}{\ifmmode~\dot{Z}\else \.{Z}\fi{}uchowski, P.~S.} \&
  \bibinfo{author}{Hutson, J.~M.}
\newblock \bibinfo{title}{Low-energy collisions of {NH}$_{3}$ and {ND}$_{3}$
  with ultracold {Rb} atoms}.
\newblock \emph{\bibinfo{journal}{Phys. Rev. A}} \textbf{\bibinfo{volume}{79}},
  \bibinfo{pages}{062708} (\bibinfo{year}{2009}).

\bibitem{campbell2009mechanism}
\bibinfo{author}{Campbell, W.} \emph{et~al.}
\newblock \bibinfo{title}{Mechanism of collisional spin relaxation in
  {$^{3}\Sigma$} molecules}.
\newblock \emph{\bibinfo{journal}{Phys. Rev. Lett.}}
  \textbf{\bibinfo{volume}{102}}, \bibinfo{pages}{13003}
  (\bibinfo{year}{2009}).

\bibitem{PhysRevLett.106.193201}
\bibinfo{author}{Parazzoli, L.~P.}, \bibinfo{author}{Fitch, N.~J.},
  \bibinfo{author}{\ifmmode~\dot{Z}\else \.{Z}\fi{}uchowski, P.~S.},
  \bibinfo{author}{Hutson, J.~M.} \& \bibinfo{author}{Lewandowski, H.~J.}
\newblock \bibinfo{title}{Large effects of electric fields on atom-molecule
  collisions at millikelvin temperatures}.
\newblock \emph{\bibinfo{journal}{Phys. Rev. Lett.}}
  \textbf{\bibinfo{volume}{106}}, \bibinfo{pages}{193201}
  (\bibinfo{year}{2011}).

\bibitem{acp-4-1461-2004}
\bibinfo{author}{Atkinson, R.} \emph{et~al.}
\newblock \bibinfo{title}{Evaluated kinetic and photochemical data for
  atmospheric chemistry: {V}olume {I} - gas phase reactions of {O}$_{x}$,
  {HO}$_{x}$, {NO}$_{x}$ and {SO}$_{x}$ species}.
\newblock \emph{\bibinfo{journal}{Atmos. Chem. Phys.}}
  \textbf{\bibinfo{volume}{4}}, \bibinfo{pages}{1461--1738}
  (\bibinfo{year}{2004}).

\bibitem{doi:10.1021/jp066359c}
\bibinfo{author}{Bahng, M.-K.} \& \bibinfo{author}{Macdonald, R.~G.}
\newblock \bibinfo{title}{Determination of the rate constant for the
  {OH}({X${^2}\Pi$)} + {OH({X}${^2}\Pi$}) $\rightarrow$ {O}({$^3$P}) +
  {H}$_2${O} reaction over the temperature range 293--373 {K}}.
\newblock \emph{\bibinfo{journal}{J. Phys. Chem. A}}
  \textbf{\bibinfo{volume}{111}}, \bibinfo{pages}{3850--3861}
  (\bibinfo{year}{2007}).

\bibitem{Avdeenkov2002}
\bibinfo{author}{Avdeenkov, A.~V.} \& \bibinfo{author}{Bohn, J.~L.}
\newblock \bibinfo{title}{Collisional dynamics of ultracold {OH} molecules in
  an electrostatic field}.
\newblock \emph{\bibinfo{journal}{Phys. Rev. A}} \textbf{\bibinfo{volume}{66}},
  \bibinfo{pages}{052718} (\bibinfo{year}{2002}).

\bibitem{Ticknor2005}
\bibinfo{author}{Ticknor, C.} \& \bibinfo{author}{Bohn, J.~L.}
\newblock \bibinfo{title}{Influence of magnetic fields on cold collisions of polar molecules}.
\newblock \emph{\bibinfo{journal}{Phys. Rev. A}} \textbf{\bibinfo{volume}{71}},
  \bibinfo{pages}{022709} (\bibinfo{year}{2005}).

\bibitem{Sawyer2008a}
\bibinfo{author}{Sawyer, B.~C.}, \bibinfo{author}{Stuhl, B.~K.},
  \bibinfo{author}{Wang, D.}, \bibinfo{author}{Yeo, M.} \& \bibinfo{author}{Ye,
  J.}
\newblock \bibinfo{title}{Molecular beam collisions with a magnetically trapped
  target}.
\newblock \emph{\bibinfo{journal}{Phys. Rev. Lett.}}
  \textbf{\bibinfo{volume}{101}}, \bibinfo{pages}{203203}
  (\bibinfo{year}{2008}).

\bibitem{Child_Book_1996}
\bibinfo{author}{Child, M.~S.}
\newblock \emph{\bibinfo{title}{Molecular collision theory}}
  (\bibinfo{publisher}{Dover Publications}, \bibinfo{year}{1996}).

\bibitem{PhysRevLett.82.3589}
\bibinfo{author}{Derevianko, A.}, \bibinfo{author}{Johnson, W.~R.},
  \bibinfo{author}{Safronova, M.~S.} \& \bibinfo{author}{Babb, J.~F.}
\newblock \bibinfo{title}{High-precision calculations of dispersion
  coefficients, static dipole polarizabilities, and atom-wall interaction
  constants for alkali-metal atoms}.
\newblock \emph{\bibinfo{journal}{Phys. Rev. Lett.}}
  \textbf{\bibinfo{volume}{82}}, \bibinfo{pages}{3589--3592}
  (\bibinfo{year}{1999}).

\bibitem{PhysRevA.74.061402}
\bibinfo{author}{Lev, B.~L.} \emph{et~al.}
\newblock \bibinfo{title}{{OH} hyperfine ground state: From precision
  measurement to molecular qubits}.
\newblock \emph{\bibinfo{journal}{Phys. Rev. A}} \textbf{\bibinfo{volume}{74}},
  \bibinfo{pages}{061402} (\bibinfo{year}{2006}).

\bibitem{PhysRevA.85.033427}
\bibinfo{author}{Stuhl, B.~K.}, \bibinfo{author}{Yeo, M.},
  \bibinfo{author}{Sawyer, B.~C.}, \bibinfo{author}{Hummon, M.~T.} \&
  \bibinfo{author}{Ye, J.}
\newblock \bibinfo{title}{Microwave state transfer and adiabatic dynamics of
  magnetically trapped polar molecules}.
\newblock \emph{\bibinfo{journal}{Phys. Rev. A}} \textbf{\bibinfo{volume}{85}},
  \bibinfo{pages}{033427} (\bibinfo{year}{2012}).

\bibitem{Bochinski2003}
\bibinfo{author}{Bochinski, J.~R.}, \bibinfo{author}{Hudson, E.~R.},
  \bibinfo{author}{Lewandowski, H.~J.}, \bibinfo{author}{Meijer, G.} \&
  \bibinfo{author}{Ye, J.}
\newblock \bibinfo{title}{Phase space manipulation of cold free radical {OH}
  molecules}.
\newblock \emph{\bibinfo{journal}{Phys. Rev. Lett.}}
  \textbf{\bibinfo{volume}{91}}, \bibinfo{pages}{243001}
  (\bibinfo{year}{2003}).

\end{thebibliography}

%% Here is the endmatter stuff: Supplementary Info, etc.
%% Use \item's to separate, default label is "Acknowledgements"

\begin{addendum}
 \item[Acknowledgments] We thank E. Cornell for useful discussions and B. Baxley for artistic contributions.
We acknowledge funding from the NSF Physics Frontier Center, DOE, AFOSR (MURI), and NIST.
\item[Contributions]B.K.S., M.T.H., M.Y., and J.Y. designed and participated in the experiment,
discussed and interpreted the results.  B.K.S. ran the day-to-day experiment and collected all
the data.  G.Q. and J.L.B. constructed the theory.  B.K.S. and J.Y. first outlined the manuscript,
and B.K.S. and G.Q. wrote the first draft. All authors discussed the results and contributed to
the preparation of the manuscript.
 \item[Competing Interests] The authors declare that they have no
competing financial interests.
 \item[Correspondence] Correspondence should be addressed to Jun Ye~(email: ye@jila.colorado.edu).
\end{addendum}

%%
%% TABLES
%%
%% If there are any tables, put them here.
%%

\end{document}